\documentclass[pre,twocolumn,preprintnumbers,amsmath,amssymb]{revtex4}
\usepackage{graphicx,epsfig}% Include figure files
\usepackage{dcolumn}% Align table columns on decimal point
\usepackage{bm}% bold math
\usepackage{psfrag}
\usepackage{plain}
\usepackage{tabularx}
\usepackage[latin1]{inputenc}
\usepackage{amsmath} 

\begin{document}
%\draft

\title{Reply to ``Comment on `Minimal size of a barchan dune' ''}
\author{E. J. R. Parteli$^1$, O. Dur\'an$^2$ and H. J. Herrmann$^{3,4}$}
\affiliation{1. Institut f\"ur Computerphysik, ICP, Universit\"at Stuttgart, Pfaffenwaldring 27, 70569 Stuttgart, Germany. \\ 2. DelftChemTech, TU Delft, Julianalaan 136, 2628 BL Delft, The Netherlands. \\ 3. Computational Physics, IfB, ETH H\"onggerberg, HIF E 12, CH-8093, Z\"urich, Switzerland. \\ 4. Departamento de F\'{\i}sica, Universidade Federal do Cear\'a - 60455-760, Fortaleza, CE, Brazil.}

\date{\today}

\begin{abstract}
We reply to the comment by Andreotti and Claudin (submitted to Phys. Rev. E, arXiv:0705.3525) on our paper ``Minimal Size of a Barchan Dune'' [Phys. Rev. E {\bf{75,}} 011301 (2007)]. We show that the equations of the dune model used in our calculations are self-consistent and effectively lead to a dependence of the minimal dune size on the wind speed through the saturation length. Furthermore, we show that Meridiani Planum ripples are probably not a good reference to estimate the grain size of Martian dune sands: the soil in the ripple troughs at the landing site is covered with nonerodible elements (``blueberries''), which increase the minimal threshold for saltation by a factor of $2.0$. We conclude that, in the absence of large fragments as the ones found at the landing site, basaltic grains of diameter $d=500 \pm 100 {\mu}$m that compose the large, typical dark Martian dunes [K. S. Edgett and P. R. Christensen, J. Geophys. Res. {\bf{96,}} 22765 (1991)] probably saltate during the strongest storms on Mars. We also show that the wind friction speed $u_{\ast} \approx 3.0$ m$/$s that we found from the calculations of Martian dunes is within the values of maximum wind speeds that occur during Martian storms a few times a decade [R. E. Arvidson {\em{et al.}}, Science {\bf{222,}} 463 (1983); H. J. Moore, J. Geophys. Res. {\bf{90,}} 163 (1985); R. Sullivan {\em{et al.}}, Nature {\bf{436,}} 58 (2005); D. J. Jerolmack {\em{et al.}}, J. Geophys. Res. {\bf{111,}} E12S02 (2006)]. In this manner, the dune model predicts that Martian dunes can be formed under present Martian conditions, with no need to assume other conditions of wind and atmosphere that could have prevailed in the past.
\end{abstract}

%\pacs{45.70.-n, 45.70.Qj}

\maketitle

In the preceeding comment \cite{Andreotti_and_Claudin_2007}, Andreotti and Claudin pretend to find inconsistencies in the dune model which has been used by Parteli {\em{et al.}} \cite{Parteli_et_al_2007} in the study of the minimal size of barchan dunes. This model, which consists of a coupled set of equations for the wind profile over the topography, the sand flux and the evolution of the topography with time, has been originally presented in refs. \cite{Sauermann_et_al_2001,Kroy_et_al_2002}; later improved in refs. \cite{Schwaemmle_and_Herrmann_2005,Duran_and_Herrmann_2006a}, and repeatedly tested through successful quantitative comparison with real wind tunnel data and with real dunes measured in the field \cite{Sauermann_et_al_2001,Sauermann_et_al_2003,Parteli_et_al_2006,Duran_and_Herrmann_2006a}. In ref. \cite{Parteli_et_al_2007}, Parteli {\em{et al.}} studied, with the dune model, the role of the wind strength and inter-dune flux for the shape and the size of the minimal dune, and used the results to obtain the wind velocity on Mars from the minimal size of Martian dunes.

The first criticism of Andreotti and Claudin \cite{Andreotti_and_Claudin_2007} is that the dune model is not self-consistent. They state that the saturation length of the sand flux, which determines the minimal dune size, should not decrease with the wind velocity because the relaxation rate is limited by the grain inertia. Next, Andreotti and Claudin \cite{Andreotti_and_Claudin_2007} find that the grain size of the ripples at Meridiani Planum landing site on Mars is $d=87 \pm 25 {\mu}$m, which is much smaller than the grain size $d=500 \pm 100$ ${\mu}$m of the larger, dark Martian dunes, as obtained from thermal inertia data \cite{Edgett_and_Christensen_1991} and used in the calculations of Parteli {\em{et al.}} \cite{Parteli_et_al_2007}. Andreotti and Claudin, then, propose an alternative explanation for the dependence of the minimal size on the wind speed: the effect of slopes. 

The comments of Andreotti and Claudin \cite{Andreotti_and_Claudin_2007} are constructive and the issues addressed by these authors deserve to be discussed in depth. We organize the present reply paper following the same structure of the preceding comment \cite{Andreotti_and_Claudin_2007}: Section I, regarding the modeling of the flux saturation length and the self-consistency of the dune model; Section II, concerning the grain size of Martian dunes sand and the reliability of the value of Martian wind velocity obtained by Parteli {\em{et al.}} \cite{Parteli_et_al_2007}; and Section III, concerning the effect of slopes on the minimal dune size.

% **********************************************************************
% **********************************************************************
% *******************  Sand transport model ****************************
% **********************************************************************
% **********************************************************************

\section{Sand transport model}

The first criticism of Andreotti and Claudin \cite{Andreotti_and_Claudin_2007} refers to an apparent inconsistency in our sand transport model. They say that, since the grain inertia is not included in the evolution of the sand flux, the saturation length determined by the ejection process can be smaller than the length needed for the grains to reach their asymptotic trajectory.

Indeed, in the current model for sand transport we assume that the characteristic length for the relaxation of the mean grain velocity in the saltation layer is much smaller than the flux relaxation length determined by grain ejection. This can lead to a discrepancy with the full model for wind shear velocities $u_*$ far from the threshold $u_{\mathrm{th}}$. In the following we calculate a modified saturation length that takes into account both processes and show that the saturation length $l_s$ is determined by the ejection process for the typical range of shear velocities found on Earth, i.e. $u_* < 3 u_{th} \approx 0.7 m/s$. 
Notice that all previous sand dunes simulation results performed with the current sand transport model are included within both ranges \cite{Parteli_et_al_2007,Sauermann_et_al_2001,Schwaemmle_and_Herrmann_2005,Kroy_et_al_2002,Duran_and_Herrmann_2006a}.

Following the original approach of Sauermann {\em et al.}~\cite{Sauermann_et_al_2001} the saltation belt is modeled as a granular fluid layer characterized by a vertically averaged mean velocity $\vec u$ and grain density $\rho$. Both magnitudes obey the mass and momentum conservation equations averaged over the $z$-axis.

\subsection{Mass conservation}

The mass conservation over the saltation layer reads \cite{Sauermann_et_al_2001}
\begin{equation}
\label{rhoxt}
\frac{\partial \rho}{\partial t} + \nabla \cdot \rho \vec u = \frac{\rho}{T_s(u)}\left(1-\frac{\rho}{\rho_s}\right)
\end{equation}
where the right hand term accounts for the interchange of particles between the saltation layer and the surface mainly due to the ejected grains by the splash. This term describes the relaxation toward saturation $\rho_s$ of the grain density $\rho$ in the saltation layer. Here, the saturation density, defined as the maximum amount of grains carried by the wind with a given shear velocity, is given by
\begin{equation}
\rho_s = \frac{2\alpha}{g}\tau_{\mathrm{th}} (U_*^2-1)
\end{equation}
and the characteristic saturation time
\begin{equation}
T_s(u) = \frac{2\alpha u}{\gamma g (U_*^2-1)}
\end{equation}
here $U_* = u_*/u_{th}$ is the relative wind shear velocity, $\tau_{\mathrm{th}}\equiv \rho_f u_{\mathrm{th}}^2$ is the threshold shear stress, $g$ is the gravity, $\gamma$ and $\alpha$ are model parameters and $\rho_f$ is the fluid density \cite{Sauermann_et_al_2001,Duran_and_Herrmann_2006a}.

\subsection{Momentum conservation}

Furthermore, the model assumes that the saltation layer over a flat surface is only subjected to a mean wind drag force and a friction force. The later accounts for the momentum lost during the inelastic grain collisions with the bed.

The momentum conservation over a flat bed is given by \cite{Sauermann_et_al_2001}
\begin{equation}
\label{uxt}
\frac{\partial \vec u}{\partial t} + \vec u \cdot \nabla \vec u = \frac{g}{u_{\mathrm{fall}}^2}|\vec v(\rho) - \vec u| (\vec v(\rho) - \vec u) - \frac{g}{2\alpha}\frac{\vec u}{u}
\end{equation}
where $u_{\mathrm{fall}}$ is the grain settling velocity. The first right-hand term represents a Newtonian drag force exerted by the wind with an effective velocity $v$, while the second gives the bed friction. 

\section{Linear analysis}

For stationary $1D$ profiles, both conservation equations (\ref{rhoxt} and \ref{uxt}) reduce to:

\begin{eqnarray}
\frac{\partial}{\partial x}(\rho u) & = & \frac{\rho}{\rho_s}\frac{\rho_s - \rho}{T(u)} \\ \nonumber
u\frac{\partial u}{\partial x} & = & \frac{g}{u_{\mathrm{fall}}^2}|v(\rho) - u|(v(\rho) - u) - \frac{g}{2\alpha}
\end{eqnarray}

This coupled system has the homogeneous solution ($\rho_s, u_s$), that correspond to the saturated state. Introducing the linear perturbations around the homogeneous solution: $\rho(x) = \rho_s (1 + \bar \rho(x))$ and $u(x) = u_s (1 + \bar u(x))$, and selecting $v(\rho_s)$ in the momentum equation, the linearized system becomes
\begin{eqnarray}
\label{sys1}
\frac{\partial \bar\rho}{\partial x} = -\frac{\bar\rho}{l_d} + \frac{\bar u}{l_v} \hspace{1cm}
\frac{\partial \bar u}{\partial x} = - \frac{\bar u}{l_v}
\end{eqnarray}
where the characteristic relaxation lengths for the mean density and velocity of the saltation layer are respectively:
\begin{eqnarray}
\label{ld}
l_d & = & u_s T_s(u_s) = \frac{2\alpha u_s^2}{\gamma g (U_*^2-1)} \\
\label{lv}
l_v & = & \frac{\sqrt{2\alpha}}{2 g}u_{\mathrm{fall}} u_s.
\end{eqnarray}
Here, the mean saturated grain velocity $u_s$ is given by
\begin{equation}
\label{us}
u_s = v_s - u_f/\sqrt{2\alpha}
\end{equation}
and the effective wind velocity in the saturated state is
\begin{equation}
\label{vs}
v_s(U_*) = \frac{u_{th}}{\kappa}\left(\ln{\frac{z_1}{z_0}} + \frac{z_1}{z_m}(U_* - 1)\right)
\end{equation}
where, for simplicity, we select the effective wind velocity at the saturated density ($v = v_s(\rho_s)$), and thus the air borne shear stress is reduced to the threshold value for sand transport~\cite{Sauermann_et_al_2001}.

The relaxation length $l_d$ accounts for the relaxation due to the ejection process, while $l_v$ includes the grain inertia, given by the falling velocity $u_{\mathrm{fall}}$, in the momentum balance.

From the first order linear system Eq.~\ref{sys1} the largest relaxation length toward saturation, defined as the saturation length $l_s$, is given by
\begin{equation}
\label{ls}
l_s = \frac{2 l_d}{(1+l_d/l_v)-|1-l_d/l_v|} = \mathrm{Max}(l_d,l_v)
\end{equation}

\begin{figure}[htpb]
\begin{center}
\includegraphics[width=1.0 \columnwidth]{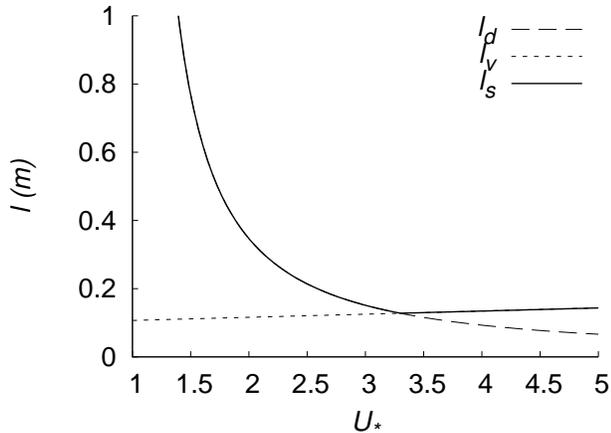}
\caption{Dependence of the density, velocity and overall relaxation lengths, $l_d$, $l_v$ and $l_s$ respectively, with the wind shear ratio (Eqs.~\ref{ld}, \ref{lv} and \ref{ls}) for Earth conditions. 
%{\bf (b)} Same for Martian conditions. All lengths are in meter.
} 
\label{fig:Lsat}
\end{center}
\end{figure}

Figure \ref{fig:Lsat} shows the saturation length $l_s$ for Earth conditions. As was pointed out by Andreotti and Claudin \cite{Andreotti_and_Claudin_2007}, the decrease of the saturation length for large shear ratios $U_*$ is clearly limited by the grain inertia. Indeed, we find that the saturation length is determined by the ejection process in the range $U_* < 3.3$, which includes most real wind conditions. Contrary to our previous assumptions \cite{Sauermann_et_al_2001,Schwaemmle_and_Herrmann_2005,Kroy_et_al_2002,Duran_and_Herrmann_2006a} and in agreement with the suggestion of Andreotti and Claudin \cite{Andreotti_and_Claudin_2007}, the spatial relaxation of the velocity of the granular layer cannot be neglected for larger shear ratios.

% **********************************************************************
% **********************************************************************
% *************** The size and density of grains on Mars ***************
% **********************************************************************
% **********************************************************************

\section{The size and density of grains on Mars}

In the comment, Andreotti and Claudin \cite{Andreotti_and_Claudin_2007} propose that the grains that constitute the sand of Martian dunes have diameter $d=87 \pm 25$ ${\mu}$m. This value has been obtained by the same authors in a previous work \cite{Claudin_and_Andreotti_2006}, in which they analysed recent photographs of Martian ripples taken by the rovers at Meridiani Planum. However, the value of grain diameter obtained by Andreotti and Claudin \cite{Andreotti_and_Claudin_2007} from the analysis of the Meridiani Planum ripples is much smaller than the grain size of the typical large intra-crater dunes as obtained from thermal inertia data, i.e. $d=500 \pm 100 {\mu}$m \cite{Edgett_and_Christensen_1991}. In fact, this value of grain diameter, which is in the range of medium to coarse sand, has been found, later, to be fairly consistent with measurements peformed by several authors in different locations on Mars \cite{Presley_and_Christensen_1997,Fenton_et_al_2003}. Andreotti and Claudin \cite{Andreotti_and_Claudin_2007}, thus, conclude that the value of grain diameter obtained from the method of thermal inertia, which has been subject of research for almost 4 decades, is wrong. 

There is no doubt that the work of Claudin and Andreotti \cite{Claudin_and_Andreotti_2006} is of relevance since the measurements of grain sizes performed by these authors are based on images of unprecedent resolution. However, care must be taken before generalizing their results of grain sizes obtained from the Meridiani Planum ripples to the typical large dark dunes on Mars.

% **********************************************************************
% ************** Threshold for saltation at Meridiani Planum ***********
% **********************************************************************

\subsection{Threshold for saltation at Meridiani Planum}

The soil of the Meridiani Planum landing site is covered with hematite spherules and fragments reaching milimeters in size. These hematite particles or ``blueberries'' are much larger and denser than the typical basaltic sand of Martian dunes. The landing site is in fact a field of {\em{coarse-grained}} ripples, whose interiors consist of fine basaltic sand in the range of $50-125$ ${\mu}$m, but which are armoured with coarse grains at their crests \cite{Sullivan_et_al_2005,Jerolmack_et_al_2006}. Hematite particles with a median diameter of about $1.0$ mm cover more than $75\%$ of the crest area of all ripples. On the other hand, the coarse-grain coverage in ripple troughs is of almost $50\%$, whereas inter-ripple areas are composed mostly of intact spherules having diameter of several milimeters, with median $3.0$ mm. In comparison, ``granule'' ripple troughs on Earth have an insignificant coverage of large fragments, the coarse particles remaining almost entirely on the ripple crests \cite{Sharp_1963,Ellwood_et_al_1975,Jerolmack_et_al_2006}.

The role of the blueberries for the transport of sand at Meridiani Planum ripple troughs is dramatic. It is well known that the presence of large particulates shielding a sand bed increases the minimal wind velocity $u_{{\ast}{\mathrm{ft}}}$ for entrainment of the finer grains into saltation \cite{Gillette_and_Stockton_1989,Nickling_and_McKenna_Neuman_1995}. Gillette and Stockton \cite{Gillette_and_Stockton_1989} found experimentally that the minimal threshold $u_{{\ast}{\mathrm{ft}}}$ of erodible grains with diameter $d=107$ ${\mu}$m increased by a factor of $k\approx2.5$ in the presence of nonerodible grains with diameters about $D=2.0-4.0$ mm having spatial coverage of $45\%$. In fact, the geometrical properties of these experiments are very similar to the ones of Meridiani Planum troughs. Indeed, a value of $k\approx 2.0$ was found later by Nickling and McKenna Neuman \cite{Nickling_and_McKenna_Neuman_1995} from experiments with larger particles, where $d = 270 {\mu}$m and $D = 18$ mm. 

On the basis of the results from the experiments mentioned in the last paragraph, it was possible to explain the formation of the Meridiani Planum ripples. As demonstrated in recent publications \cite{Sullivan_et_al_2005,Jerolmack_et_al_2006}, there are strong evidences that the minimal wind velocity required to mobilize the sand grains at Meridiani Planum ripple troughs has been effectively increased by a factor $k$ of about $2.0-2.5$, as observed in experiments with sand bed shielded by nonerodible roughness mentioned above. 

\begin{figure}[htpb]
\begin{center}
\includegraphics[width=1.0\columnwidth]{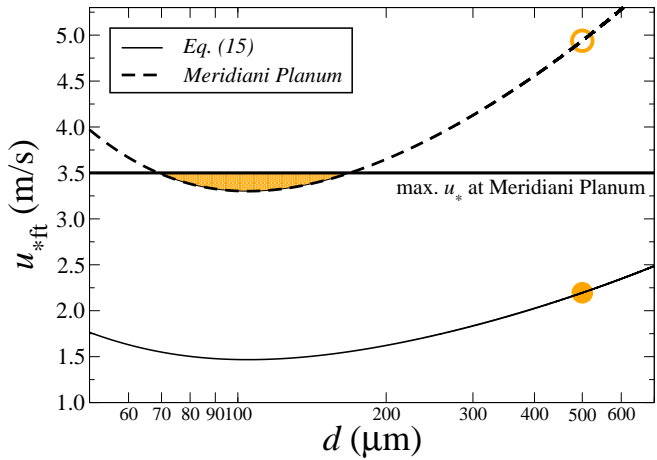}
\caption{Threshold wind shear velocity $u_{{\ast}{\mathrm{ft}}}$ for direct particle entrainment into saltation on Mars. The solid curve shows $u_{{\ast}{\mathrm{ft}}}$ calculated with Eq. (\ref{eq:u_ft}). At the Meridiani Planum landing site, the presence of the blueberries increases $u_{{\ast}{\mathrm{ft}}}$ of Martian sand by a factor of $2.25$ (dashed line). The straight line shows the maximum peak of shear velocity, $u_{\ast} = 3.5$ m$/$s, associated with the largest storm at the landing site. During such a storm, only particles between $69$ and $168$ ${\mu}$m are expected to be entrained by wind at the landing site (dashed area). The threshold for the grain diameter of the dark Martian sand dunes, $d=500$ ${\mu}$m, calculated with Eq. (\ref{eq:u_ft}), is indicated by the filled circle. The empty circle shows $u_{{\ast}{\mathrm{ft}}}$ for $d=500$ ${\mu}$m at the Meridiani Planum landing site: such coarse grains cannot saltate at the landing site, at present conditions.} 
\label{fig:meridiani}
\end{center}
\end{figure}

In the absence of nonerodible large fragments, the minimal wind velocity required to entrain sand grains into saltation can be calculated  with the equation \cite{Iversen_and_White_1982}: 
\begin{equation}
u_{{\ast}{\mathrm{ft}}}= A\,\sqrt{{\frac{{({\rho}_{\mathrm{grain}}-{\rho}_{\mathrm{fluid}})}gd}{{\rho}_{\mathrm{fluid}}}}}, \label{eq:u_ft}
\end{equation}
where $g$ is gravity and ${\rho}_{\mathrm{fluid}}$ and ${\rho}_{\mathrm{grain}}$ are the densities of the air, respectively of the grains. The Shields parameter $A$ is given by \cite{Iversen_and_White_1982}:
\begin{equation}
A = 0.129 {\left[{{\frac{{\left({1 + 6.0 \times 10^{-7}/{{{\rho}_{\mathrm{grain}}}gd^{2.5}}}\right)}^{0.5}}{{\left({1.928{\mbox{Re}}_{{\ast}{\mathrm{ft}}}^{0.092}}-1\right)}^{0.5}}}}\right]} \label{eq:Shields_parameter_a}
\end{equation} 
for $0.03 \leq {\mbox{Re}}_{{\ast}{\mathrm{ft}}} \leq 10$ and 
\begin{eqnarray}
A  && = 0.129 {\left({1 + 6.0 \times 10^{-7}/{{{\rho}_{\mathrm{grain}}}gd^{2.5}}}\right)}^{0.5} \nonumber \\ & & \cdot {\left\{{1 - 0.0858 \exp{\left[{-0.0617({\mbox{Re}}_{{\ast}{\mathrm{ft}}}-10)}\right]}}\right\}}  \label{eq:Shields_parameter_b}
\end{eqnarray}
for ${\mbox{Re}}_{{\ast}{\mathrm{ft}}} \geq 10$, where $\nu$ is the kinematic viscosity, ${\mbox{Re}}_{{\ast}{\mathrm{ft}}}$ is the friction Reynolds number ${\mbox{Re}}_{{\ast}{\mathrm{ft}}} \equiv u_{{\ast}{\mathrm{ft}}}d/{\nu}$, and the constant $6.0 \times 10^{-7}$ has units of ${\mbox{kg}}{\cdot}{\mbox{m}}^{0.5}{\cdot}{\mbox{s}}^{-2}$, while all other numbers are dimensionless. The solid curve in fig. \ref{fig:meridiani} shows $u_{{\ast}{\mathrm{ft}}}$ as function of the grain diameter calculated with Eq. (\ref{eq:u_ft}) using ${\rho}_{\mathrm{grain}} = 3200$ kg$/$m$^3$, ${\rho}_{\mathrm{fluid}} =0.02$ kg$/$m$^3$, and ${\nu} = 6.35 \cdot 10^{-4}$ m$^2/$s \cite{Jerolmack_et_al_2006}.

We follow the idea of Jerolmack {\em{et al.}} \cite{Jerolmack_et_al_2006} and calculate the modified threshold for saltation at Meridiani Planum, $k\,u_{{\ast}{\mathrm{ft}}}$, taking the average value $k=2.25$. The result is shown by the dashed curve in fig. \ref{fig:meridiani}. In this figure, the full, straight line represents the maximum allowed wind friction speed during the gusts of dust storm at Meridiani Planum: $u_{\ast}=3.5$ m$/$s. This value of wind speed, which is probably achieved once in intervals of years \cite{Sullivan_et_al_2005}, is estimated to be an upper bound because larger wind speeds would result in saltation of the hematite spherules, which evidently did not occur during formation of the ripples. As explained previously, the winds that formed the ripples at Meridiani Planum landing site have friction speed $u_{\ast}$ in the range $2.5-3.5$ m$/$s, the lower bound corresponding to the minimal threshold for creeping motion of the hematite grains \cite{Jerolmack_et_al_2006}.

Although the estimation of the modified threshold for saltation (dashed line of fig. \ref{fig:meridiani}) is very crude, it suggests that the wind strength that sculped the soils of Meridiani Planum was just sufficient to entrain the grains of smallest saltation threshold values, as recognized in ref. \cite{Claudin_and_Andreotti_2006}. The dashed area of fig. \ref{fig:meridiani} corresponds to the range of grain sizes that are entrained by the wind into saltation at Meridiani Planum, $69 \leq d \leq 168$ ${\mu}$m, assuming $u_{{\ast}{\mathrm{ft}}}$ is about $2.25$ times the value calculated with Eq. (\ref{eq:u_ft}). The minimum for saltation occurs in fact at about $100$ ${\mu}$m, which is well within the range of grain sizes of the sand found in the interior of coarse-grained ripples; on the matrix bed in the ripple troughs, and within small pits and craters at Meridiani Planum, which apparently serve as particle traps \cite{Sullivan_et_al_2005}. 

For illustration, the value of $u_{{\ast}{\mathrm{ft}}}$ obtained with Eq. (\ref{eq:u_ft}) for the grain diameter of Martian dunes, $d=500{\mu}$m, is shown by the filled circle in fig. \ref{fig:meridiani}. The empty circle shows the modified threshold, $2.25u_{{\ast}{\mathrm{ft}}}$, for $d=500{\mu}$m. We see that saltation of such coarse grains at Meridiani Planum landing site would require a wind of $u_{\ast}\approx 5.0$ m$/$s, which is much larger than the maximum value, $3.5$ m$/$s. However, it is clear from fig. \ref{fig:meridiani} that basaltic grains much larger than those of the landing site can be entrained by a wind of strength $u_{\ast}=3.5$ m$/$s, in places where sand is not shielded by nonerodible elements.

In conclusion, the threshold for saltation transport at Meridiani Planum landing site is modified due to the presence of nonerodible hematite fragments on the soil. Thus, provided other factors as sand induration \cite{Schatz_et_al_2006} are not affecting the local threshold for saltation, it is very plausible that the grains of Martian dunes, which have diameter $d=500 \pm 100 {\mu}$m \cite{Edgett_and_Christensen_1991}, are effectively entrained by formative winds of strength $2.5 \leq u_{\ast} \leq 3.5$ m$/$s under present Martian conditions, since the threshold for entrainment of such coarse grains is exceeded at such values of $u_{\ast}$.

% **********************************************************************
% ************** The wind velocity that forms dunes on Mars ************
% **********************************************************************

\subsection{The wind velocity that forms dunes on Mars}

The main criticism in the comment by Andreotti and Claudin regarding the results on Martian dunes is that the dune model predicts that ``very strong'' winds \cite{Andreotti_and_Claudin_2007} are required to form the Martian dunes. We recall the value of Martian wind shear velocity obtained in Parteli {\em{et al.}} from the minimal dune size \cite{Parteli_et_al_2007}: $u_{\ast}\approx 3.0$ m$/$s. However, values of $u_{\ast}$ about $3.0$ m$/$s are within maximum values of shear velocity on Mars, and occur only during the strongest dust storms \cite{Arvidson_et_al_1983}. Sand transport on Mars is, thus, expected to consist of short duration events (a few minutes) a few times a decade \cite{Moore_1985,Sullivan_et_al_2005}, and does not occur under typical Martian wind velocities, which are between $0.3$ and $0.7$ m$/$s \cite{Sutton_et_al_1978}. We conclude that the value $u_{\ast} \approx 3.0$ m$/$s found in Parteli {\em{et al.}} \cite{Parteli_et_al_2007} from the shape of Martian sand dunes is consistent with real values of wind velocities expected to occur during sand transport on Mars.

Furthermore, as it will be shown in a future publication \cite{Parteli_and_Herrmann_2007}, Martian dunes of different shapes and sizes and at different locations on Mars can be explained without necessity to assume that they were formed ``in the past under very strong winds'' as stated in ref. \cite{Andreotti_and_Claudin_2007}. The calculations using the model presented in Parteli {\em{et al.}} \cite{Parteli_et_al_2007} show that the wind velocity on Mars in fact does not exceed $3.0$ m$/$s. 

Indeed, the differences in minimal dune sizes on Mars can be explained on the basis of different local conditions: the average density of the Martian air may vary by a factor of $2.0$ depending on the location, and in this manner, in places where the air is denser (e.g. on the north pole), the minimal dune width can be correspondingly smaller under the same wind $u_{\ast}=3.0$ m$/$s. This shear velocity is between $1.0$ and $2.0$ times the threshold for saltation on Mars, similarly to the situation on Earth, and is of the same order of magnitude as real Martian sand-moving winds \cite{Arvidson_et_al_1983,Moore_1985,Sullivan_et_al_2005,Jerolmack_et_al_2006}.

% **********************************************************************
% **********************************************************************
% *********************** Linear stability analysis ********************
% **********************************************************************
% **********************************************************************

\section{Linear stability analysis: relation between the unstable dune wavelength and the saturation length}

In their comment on ``Minimal size of barchan dunes", Andreotti and Claudin \cite{Andreotti_and_Claudin_2007} also proposed a novel mechanism to understand the apparent scaling of the minimal dune size with the inverse of the wind shear stress, besides the scaling of the saturation length which arises from the derivation of the dune model (Section I). 

Following the work of Rasmussen {\em et al.} \cite{Rasmussen_1996}, 
Andreotti and Claudin include the dependence of the threshold shear stress $u^2_{th}$ on the local slope $\tan{\alpha}\equiv\partial_x h$, given originally by $u^2_{th} (\cos{\alpha} + \sin{\alpha}/\tan\theta)$ \cite{Rasmussen_1996}, where $\theta$ is the angle of repose, into the linear stability analysis of the equations for the dune evolution. However, for small slopes they approximate the above expression by $u^2_{th} (1+\partial_x h/\tan\theta)$ instead of the more appropriate one: $u^2_{th} (1+(1-\tan\theta)\partial_x h/\tan\theta)$. Furthermore, the scaling of the saturated flux with the local slope is not only reduced to the threshold shear stress. In fact, it is more complex and can be written as \cite{Rasmussen_1996}
\begin{equation}
\label{qs}
q_s = \chi (u_*^2 - u_{th}^2 M) u_{th}/\sqrt{M}
\end{equation}
where $M = 1+\partial_x h/a$, the constant $a = \tan\theta/(1-\tan\theta)$ and $u_*$ is the wind shear velocity. 
%For a repose angle $\theta = 34^o$, the scaling term $M$ can be approximated by $1 + \partial_x h/2$.

With this slope dependence the linear perturbation of the normalized saturated flux in the Fourier space can be written as:
\begin{equation}
\hat{q_{\mathrm{s}}} = \frac{1}{U^2_*-1}\left(U^2_* (A + i B) - \frac{i}{2 a}(U^2_*+1)\right) k \hat h,
\end{equation}
where $U_* \equiv u_*/u_{th}$ and $\hat h$ is the surface Fourier transform. 

The linear stability analysis shows that the surface perturbation $\hat h$ is unstable \cite{Andreotti_et_al_2002} and has a growth rate $\sigma$:
\begin{equation}
\sigma = \frac{\chi U^2_*k^2}{1+k^2 l^2_s(U_*)}\left(B - \frac{U^2_*+1}{2 a U^2_*} - A k {\ell}_{\mathrm{s}}\right)
\end{equation}
and thus, the marginally stable wavelength $\lambda_c$, at which $\sigma = 0$, is:
\begin{equation}
\lambda_c = \frac{4\pi a A U^2_*}{U^2_*(2 a B-1)-1} {\ell}_{\mathrm{s}}(U_*).
\end{equation}
%Here, the term $2 a B - 1$ accounts for the increase of the transport threshold for positive slopes, and the reduction of the mean grain velocity due to gravity also for positive slopes, given by the term $1/\sqrt{M}$ in Eq.~\ref{qs}. 
For a repose angle $\theta\approx 34^o$ ($a\approx 2$) and the typical values $B = 1.5$ and $A = 5$ \cite{Andreotti_and_Claudin_2007}, $\lambda_c$ becomes:
\begin{equation}
\lambda_c = \frac{8\pi A U^2_*}{5 U^2_*-1} {\ell}_{\mathrm{s}}(U_*) \approx 8\pi {\ell}_{\mathrm{s}}(U_*) ,
\end{equation}
where, due to the factor $2 a B -1 \gg 1$, it is clear that the main dependence on the shear stress {\em{is}} in the saturation length. Notice that if $a$ were simply $\tan\theta$ as suggested in \cite{Andreotti_and_Claudin_2007} the constant $2 a B -1 \approx 1$.

Summarizing, in the selection of the dominant dune wavelength, we find that the ``second order" effects in the saturated flux apparently counteracts each other. Therefore, the contribution of wind speed to the dune size selection is mainly through the scaling of the saturation length, with no simple alternatives.

{\bf{Acknowledgements}} --- This work was partially supported by Volkswagenstiftung, DFG and the Max-Planck Prize. EJRP acknowledges support from CAPES, Bras\'{\i}lia/Brazil.


\begin{thebibliography}{99}

\bibitem{Andreotti_and_Claudin_2007} 
B. Andreotti and P. Claudin, preceeding comment, submitted to Phys. Rev. E (2007), arXiv:0705.3525.

\bibitem{Parteli_et_al_2007} E. J. R. Parteli, O. Dur\'an and H. J. Herrmann, Phys. Rev. E {\bf{75,}} 011301 (2007).

\bibitem{Sauermann_et_al_2001} G. Sauermann, K. Kroy and H. J. Herrmann, Phys. Rev. E {\bf{64,}} 31305 (2001).

\bibitem{Kroy_et_al_2002} K. Kroy, G. Sauermann and H. J. Herrmann, Phys. Rev. E {\bf{66,}} 031302 (2002).

\bibitem{Schwaemmle_and_Herrmann_2005} V. Schw\"ammle and H. J. Herrmann, The European Physical Journal E {\bf{16,}} 57 (2005).

\bibitem{Duran_and_Herrmann_2006a}
O. Dur\'an and H. J. Herrmann, JSTAT {\bf{P07011}} (2006).

\bibitem{Sauermann_et_al_2003}
G. Sauermann, J. S. Andrade Jr., L. P. Maia, U. M. S. Costa, A. D. Ara\'ujo and H. J. Herrmann, Geomorphology {\bf{54,}} 245 (2003).

\bibitem{Parteli_et_al_2006}
E. J. R. Parteli, V. Schw\"ammle, H. J. Herrmann, L. H. U. Monteiro and L. P. Maia, Geomorphology {\bf{81,}} 29 (2006).

\bibitem{Edgett_and_Christensen_1991} K. S. Edgett and P. R. Christensen, Journal of Geophysical Research {\bf{96}}, 22765 (1991).

\bibitem{Anderson_and_Haff_1991} 
R. S. Anderson and P. K. Haff Acta Mechanica (Suppl.) {\bf{1}}, 21 (1991).

\bibitem{Ungar_and_Haff_1987} J. E. Ungar and P. K. Haff, Sedimentology {\bf{34,}} 289 (1987).

\bibitem{Owen_1964} P. R. Owen, Journal of Fluid Mechanics {\bf{20}}, 225 (1964).

\bibitem{Andreotti_2004} B. Andreotti, J. Fluid Mech. {\bf{510,}} 47 (2004).

\bibitem{Andreotti_et_al_2002}
B. Andreotti, P. Claudin and S. Douady, Eur. Phys. J. B. {\bf{28,}} 321 (2002).

\bibitem{Claudin_and_Andreotti_2006} P. Claudin and B. Andreotti, Earth and Planetary Science Letters {\bf{252}}, 30 (2006).

\bibitem{Presley_and_Christensen_1997} M. Presley and P. R. Christensen, Journal of Geophysical Research {\bf{102}}, 6551 (1997).

\bibitem{Fenton_et_al_2003} L. K. Fenton, J. L. Bandfield, and A. W. Wartd, Journal of Geophysical Research {\bf{108}}(E12), 5129, doi:10.1029$/$2002JE002015 (2003).

\bibitem{Sullivan_et_al_2005}
R. Sullivan, {\em{et al.}}, Nature {\bf{436}}, 58 (2005).

\bibitem{Jerolmack_et_al_2006}
D. J. Jerolmack, D. Mohrig, J. P. Grotzinger, D. A. Fike and W. A. Watters, Journal of Geophysical Research {\bf{111,}} E12S02, doi:10.1029$/$2005JE002544 (2006).

\bibitem{Sharp_1963}
R. P. Sharp, J. Geol. {\bf{71,}} 617 (1963).

\bibitem{Ellwood_et_al_1975}
J. M. Ellwood, P. D. Evans, and I. G. Wilson {\bf{45,}} 554 (1975).

\bibitem{Gillette_and_Stockton_1989}
D. Gillette and P. H. Stockton, J. Geophys. Res. {\bf{94,}} 12885 (1989).

\bibitem{Nickling_and_McKenna_Neuman_1995}
W. G. Nickling and C. McKenna Neuman, Sedimentology {\bf{50,}} 247--263 (1995).

\bibitem{Iversen_and_White_1982}
J. D. Iversen and B. R. White, Sedimentology {\bf{29,}} 111 (1982).

\bibitem{Schatz_et_al_2006}
V. Schatz, H. Tsoar, K. S. Edgett, E. J. R. Parteli and H. J. Herrmann, J. Geophys. Res. {\bf{111}}(E4), E04006, doi:10.1029/2005JE002514 (2006).

\bibitem{Arvidson_et_al_1983}
R. E. Arvidson, E. A. Guinness, H. J. Moore, J. Tillmann, and S. D. Wall, Science {\bf{222}}, 463 (1983).

\bibitem{Moore_1985}
H. J. Moore, Journal of Geophysical Research {\bf{90}}, 163 (1985).

\bibitem{Sutton_et_al_1978} J. L. Sutton, C. B. Leovy, and J. E. Tillman, Journal of the Atmospheric Sciences {\bf{35}}, 2346 (1978).

\bibitem{Kroy_et_al_2005}
K. Kroy, S. Fischer, and B. Obermayer, Journal of Physics: Condensed Matter {\bf{17}}, S1299 (2005).

\bibitem{Parteli_and_Herrmann_2007} 
E. J. R. Parteli and H. J. Herrmann, submitted to Phys. Rev. E (2007), arXiv:0705.0809.

\bibitem{Rasmussen_1996}
K. R. Rasmussen, J. D. Iversen and P. Rautaheimo, Geomorphology {\bf 17}, 19 (1996).

\end{thebibliography}
\end{document}